\pdfoutput=1
\documentclass[preprint]{aastex62}

\usepackage{natbib}
\received{April 2, 2019}
\revised{April 26, 2019}
\accepted{May 1, 2019}
\published{May 20, 2019}
\submitjournal{ApJL}

\shorttitle{The Physcial Nature of Spiral Wave Patterns}
\shortauthors{Kang et al.}

\begin{document}

\title{The Physical Nature of Spiral Wave Patterns in Sunspots}

\author[0000-0003-3540-4112]{Juhyung Kang}
\affil{Astronomy Program, Department of Physics and Astronomy, Seoul National University, Seoul 08826, Republic of Korea}

\author[0000-0002-7073-868X]{Jongchul Chae}
\affil{Astronomy Program, Department of Physics and Astronomy, Seoul National University, Seoul 08826, Republic of Korea}

\author[0000-0001-6423-8286]{Valery M. Nakariakov}
\affil{Centre for Fusion, Space and Astrophysics, Physics Department, University of Warwick, Coventry CV4 7AL, UK}
\affil{School of Space Research, Kyung Hee University, Yongin 17104, Republic of Korea}

\author[0000-0001-7460-725X]{Kyuhyoun Cho}
\affil{Astronomy Program, Department of Physics and Astronomy, Seoul National University, Seoul 08826, Republic of Korea}

\author[0000-0001-8619-9345]{Hannah Kwak}
\affil{Astronomy Program, Department of Physics and Astronomy, Seoul National University, Seoul 08826, Republic of Korea}

\author[0000-0001-9455-3615]{Kyeore Lee}
\affil{Astronomy Program, Department of Physics and Astronomy, Seoul National University, Seoul 08826, Republic of Korea}

\correspondingauthor{Jongchul Chae}
\email{jcchae@snu.ac.kr}

\begin{abstract}

Recently, spiral wave patterns (SWPs) have been detected in 3 minute oscillations of sunspot umbrae, but the nature of this phenomenon has remained elusive.
We present a theoretical model that interprets the observed SWPs as the superposition of two different azimuthal modes of slow magnetoacoustic waves driven below the surface in an untwisted and non-rotating magnetic cylinder. We apply this model to SWPs of the line-of-sight (LOS) velocity in a pore observed by the Fast Imaging Solar Spectrograph installed at the 1.6~m Goode Solar Telescope.
One- and two-armed SWPs were identified in instantaneous amplitudes of LOS Doppler velocity maps of 3 minute oscillations.
The associated oscillation periods are about 160 s, and the durations are about 5 minutes.
In our theoretical model, the observed spiral structures are explained by the superposition of non-zero azimuthal modes driven 1600~km below the photosphere in the pore.
The one-armed SWP is produced by the slow-body sausage ($m=0$) and kink ($m=1$) modes, and the two-armed SWP is formed by the slow-body sausage ($m=0$) and fluting ($m=2$) modes of the magnetic flux tube forming the pore.

\end{abstract}


\section{Introduction} \label{sec:intro}

Wave motions are a conspicuous dynamic phenomenon observed in sunspots.
The first detection of sunspot waves in the chromosphere was reported by \citet{Beckers1969}.
Subsequent works revealed that the predominant period of the waves is 5 minutes in the umbral photosphere \citep{Bhatanagar1972}, and 3 minutes in the chromosphere \citep{Beckers1972}.
Sunspot waves were also observed in the transition region and corona with the periods of less than three minutes \citep[e.g.][]{DeMoortel2002, Sych2009, Tian2014}.
Furthermore, a radially propagating wave pattern was detected in the sunspot penumbra that is known as running penumbral waves \citep[RPWs;][]{Giovanelli1972,Zirin1972}.
A comprehensive review of sunspot waves can be found in \citet{Khomenko2015}.

The nature of 3 minute chromospheric oscillations has been attributed to upward propagating slow magnetoacoustic waves \citep{Lites1984,Centeno2006}.
\citet{Centeno2006} clearly showed the propagating property of the waves by measuring the phase difference between the time series of the line-of-sight (LOS) velocity in the photosphere and that in the chromosphere.
In the same context, the RPWs have been interpreted as the slow waves propagating along the inclined magnetic field lines \citep{Bloomfield2007,Lohner2015}.

The plausible driving sources of sunspot waves are external $p$-modes and internal magnetoconvection.
The external driving scenario assumes that $f$- and $p$-mode waves in a quiet Sun propagate into a sunspot. A fraction of the energy of the incident $f$- and $p$-mode is absorbed by its conversion into a slow magnetoacoustic mode at the plasma-$\beta$ equal to one layer \citep[e.g.,][]{Cally1994,Cally1997,Cally2003}.
\citet{Zhao2013} successfully observed the absorption of the $f$- and $p$-mode wave energy in a sunspot in the $k-\omega$ diagram.
In the internal driving model, magnetoconvection occurring inside a sunspot can excite the waves.
The radiative magnetohydrodynamics simulations of the magnetoconvection showed that multi-frequency waves can be generated in a magnetic concentration region such as a sunspot \citep{Jacoutot2008}.
\citet{Chae2017} found that the wave energy flux was enhanced around the light bridge and umbral dots, and they concluded that the magnetoconvection may be the driving source of 3 minute oscillations.
The internal excitation was further supported by \citet{Cho2019}'s identification of several patterns characterized by oscillation centers and radial propagation above individual umbral dots that are under substantial changes.
Recent works suggested that an internal driving source may be located, below the sunspot photosphere down to 5~Mm in the sunspot's flux tube, by analyzing the photospheric fast-moving wave patterns \citep{Zhao2015,Felipe2017}.

Interestingly, recent observational works reported that in the horizontal plane, 3 minute oscillations often appear in sunspot umbrae as one- and two-armed spiral wave patterns \citep[SWPs;][]{Sych2014, Su2016, Felipe2019}.
SWPs apparently propagate radially out at the velocity of around 20~km~s$^{-1}$, and also propagate upward \citep{Su2016}.
Because these propagating properties are similar to RPWs, \citet{Su2016} concluded that observed SWPs could be associated with the slow waves propagating along a twisted magnetic field.
\citet{Sych2014}, however, pointed out that the magnetic field should be uniformly twisted in low-$\beta$ plasma of sunspots, and it cannot contribute to the non-uniformity of a SWP.
Moreover, the observed SWPs highlight the structure of the wavefront in a certain horizontal cross section of the magnetic flux tube, which does not require the flux tube twisting.
Very recently, \citet{Felipe2019} also concluded that although the twist can affect the shape of the observed SWPs, it is not their main cause.

In this Letter we present a simple model that SWPs can naturally appear in an untwisted magnetic flux tube when non-axisymmetric disturbances from below the surface are taken into account. We observationally identify one- and two-armed SWPs in a pore in Doppler velocity maps of the H$\alpha$ line profiles, and develop a theoretical model explaining the appearance of SWPs.
In section \ref{sec:obs}, we describe the observations, and summarize observational results.
In section \ref{sec:model} we describe the theoretical model that reproduces the SWPs, together with their simulation.
Finally, in Section \ref{sec:disc} we discuss and conclude the main results.

\section{Observation} \label{sec:obs}

We observed a pore in NOAA 12078 on 2014 June 3 from 16:48:41 to 17:56:32~UT with the 1.6~m Goode Solar Telescope.
The target was located at $x=160\arcsec$, $y=-300\arcsec$ when we started the observation.
In this study, we used the data acquired by the Fast Imaging Solar Spectrograph (FISS) in the H$\alpha$ band, and this is the same data analyzed previously in \citet{Chae2015}.
The FISS scanned the pore with a spectral sampling of 0.019~\AA and spatial sampling of $0.\arcsec16$, covering a field of view of $20\arcsec$ by $40\arcsec$.
The exposure time was 30~ms, and the time cadence of the data was 20 s.
The basic calibration was performed as described by \citet{Chae2013}.
We measured the LOS Doppler velocities for all data pixels by using the lambdameter method \citep{Chae2013-2} with the lambdameter chord of 0.4~\AA.
To highlight 3 minute oscillations, we filtered the data in frequency, leaving only the frequencies of $5.5-9$~mHz.

\begin{figure*}
	\plotone{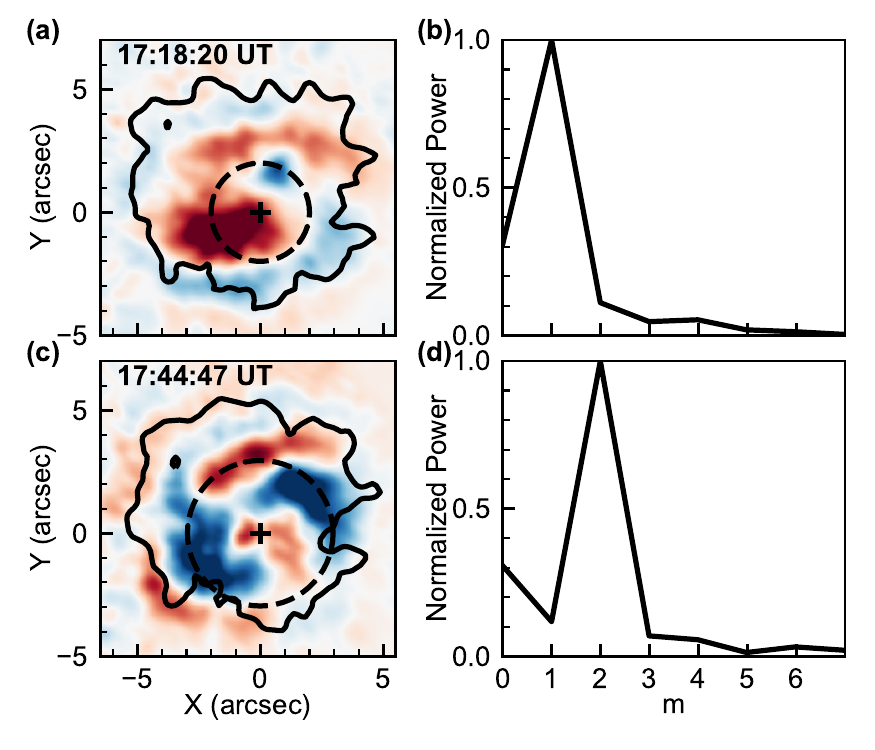} 
	\caption{
		Snapshots of the LOS Doppler velocity maps (left panels), and their time-averaged azimuthal power spectra in the azimuthal direction along the dashed line (right panels).
		Blue (red) color represents upflows (downflows), and the saturation amplitude of velocity is 3~km~s$^{-1}$.
		The black contour represents the boundary of the pore.
		The cross symbol indicates the center of the dashed line, and this position is set to be the origin.
		The radius of the dashed line is 2\arcsec\ for the one-armed SWP (a) and 3\arcsec\ for the two-armed SWP (c).
	}
	\label{fig:snapshot}
\end{figure*}

From the filtered Doppler velocity maps, we identified three SWPs, but here we deal only with the the case studies of one- and two-armed SWPs.
The left panels of Figure \ref{fig:snapshot} show the one- and two-armed SWPs measured from the velocity maps at 17:18:20 UT and 17:44:47 UT, respectively.
The time evolution of these patterns during one cycle is illustrated in Figure~\ref{fig:arm_1} and Figure~\ref{fig:arm_2}, and associated animations are available online.
These wave patterns rotated in the counterclockwise direction.
The spiral arm structures are seen to move outward, and their amplitude become to zero near the boundary of the pore.
On the other hand, the center of the arms moved abruptly inward direction while rotating, like a spiral; hereafter, we call this as \textit{spiraling}.
We determined the duration of the SWPs by the visual inspection of the rotating motion. It was found to be about 4 minutes for the one-armed spiral, and 5 minutes for the two-armed spiral.
From the wavelet analysis, we estimated the oscillation period of SWPs at about 120 s at the center of the pore and at about 250 s near its boundary.
The period averaged over the pore is about 165 s.

To identify the spatial fluctuations of the patterns in the azimuthal direction, the discrete Fourier transform was applied along the dashed line.
The Figure \ref{fig:snapshot} shows the time-averaged azimuthal power spectra of the two SWPs constructed along the two circles marked by the dashed curves.
At these two radii, the power of non-zero azimuthal mode $m$ is the largest.
In the case of the one-armed SWP, most of the power is concentrated at $m=0$ and $m=1$ (panel (b)).
For the two-armed spiral, the power is concentrated at the $m=0$ and $m=2$ (panel (d)).
These indicate that the SWPs are composed of at least two azimuthal modes.
We found that during each event, both the azimuthally symmetric modes ($m=0$) and the non-symmetric mode ($m=1$ or 2) appeared and disappeared together.
The power of $m=0$ mode at the chosen radius fluctuated substantially for the period of about 80 s, whereas the power of $m=1$ or 2 mode changed slowly with time.

We detected such SWPs in other sunspots as well.
Roughly speaking, from an one hour observation, two or three SWPs occurred inside each sunspot.
The rotation direction of the SWPs did not have any hemispheric dependence.
In some cases, in fact, two SWPs of opposite rotation directions were observed in the same sunspot at two different times.
Even though such SWPs were detected in any types of sunspots, the spiral arms were simply shaped in small axisymmetric sunspots.
The details of these observational results will be described in a subsequent paper.

\section{Modeling} \label{sec:model}

To interpret the detected SWPs, we first consider azimuthal wave modes in an untwisted uniform thick magnetic cylinder with the magnetic field along the $z$ direction, following \citet{Edwin1983}.
The observed pore is well compatible with this assumption because it contains a straight field that is confined to the pore's boundary.
The internally oscillatory solution (body waves) of the transverse and longitudinal velocity components in cylindrical coordinates ($r,\theta,z$) are given as follows \citep{Spruit1982, Lopez2016}:
\begin{eqnarray}
	v_r&=&-\frac{\omega^2-k^2 c_s^2}{\omega^2 n^2} A_m J'_m(n r) \exp{i \left(kz+m\theta-\omega t\right)} , \label{eq:vr} \\
	v_z&=&-i \frac{k c_s^2}{\omega^2} A_m J_m(nr) \exp{i \left(kz+m\theta-\omega t\right)}, \label{eq:vz}
\end{eqnarray}
where $k$ is the wavenumber along the field, $\omega$ is the frequency, $c_s$ is the sound speed, $A_m$ is the amplitude of an azimuthal mode $m$, $J_m$ is the Bessel function of the first kind, and $J'_m$ is its derivative.
In this study, we follow the general naming convention for the integer azimuthal modes: \textit{sausage} mode for $m=0$, \textit{kink} mode for $m=1$, and \textit{fluting} modes for $m\ge2$.

The effective radial wavenumber $n$ is given by \citep{Edwin1983}
\begin{equation}
	n^2=\frac{\left(\omega^2-c_s^2 k^2\right) \left(\omega^2-c_A^2 k^2\right)}{\left(c_s^2+c_A^2\right)\left(\omega^2-c_T^2 k^2\right)} ,
\end{equation}
where $c_A$ is the Alfv\'en speed, and $c_T$ is the tube speed, $c_T^2=c_s^2 c_A^2 /(c_s^2+c_A^2)$.
For body waves $n^2$ must be positive, and for slow modes the phase speed $\omega/k$ lies between the tube speed and sound speed  \citep{Roberts2006}.

\begin{figure*}
	\plotone{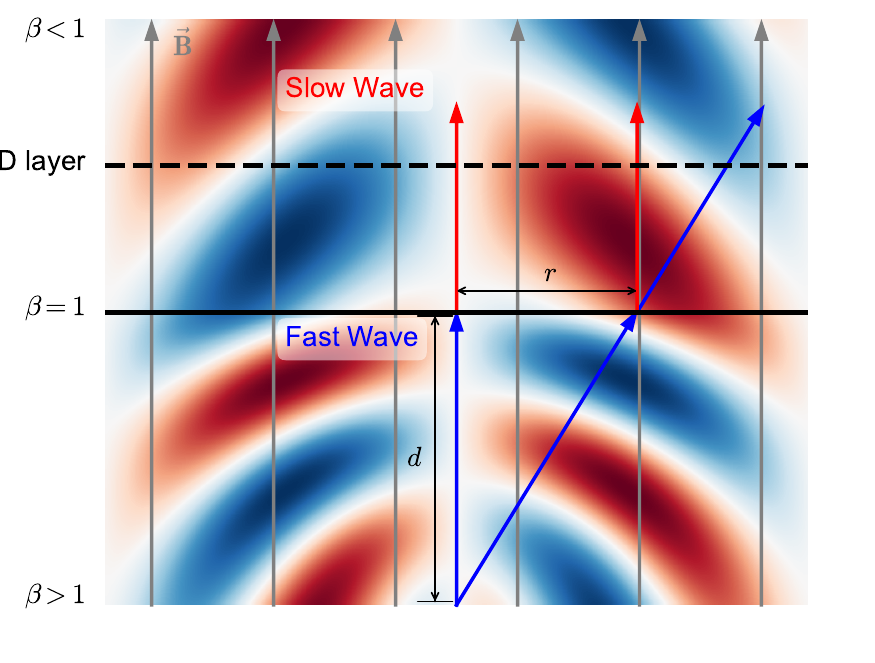}
	\caption{Schematic images of the longitudinal velocities $v_z$ in the $m=1$ mode in the $x-z$ plane.
		The driving source of the wave is located at the center of the bottom.
		Blue (red) color represents the upflows (downflows).
		The black solid line indicates the $\beta=1$ layer and the dashed line denotes the detection layer (D layer).
		Magnetic field lines are shown by the gray arrows.
		The propagating direction of the fast (slow) wave is shown by the blue (red) arrow.
	}
	\label{fig:model}
\end{figure*}

In addition, we assume that the driving source of the wave is located below the photosphere inside the flux tube.
This approach is in line with the suggestion of \citet{Zhao2015} and \cite{Felipe2017} made to interpret the photospheric fast-moving radial wave patterns.
In this scenario a fast mode wave is driven at the high-$\beta$ region, then it propagates quasi-isotropically to the $\beta=1$ layer (see Figure \ref{fig:model}).
Thus, the arrival time $t_A(r)$ at the $\beta=1$ layer is given as a function of the transverse distance $r$ from the center of the source, \begin{equation}
	t_A(r)=\frac{\sqrt{r^2+d^2}}{v_\mathrm{fast}} ,
\end{equation}
where $d$ is the depth of the source and $v_\mathrm{fast}$ is the averaged propagation speed of the fast wave in the high-$\beta$ region.
For simplicity, here we have assumed the constancy of the propagation speed and neglected the effect of refraction and reflection.
After arriving at the $\beta=1$ layer, a portion of the fast wave is converted to the slow wave \citep{Cally2001} which then propagates along the field.
For that reason, we can observe the radially propagating wave patterns when the slow mode reaches the detection layer.
With the use of this effect, we can re-write the Equation (\ref{eq:vz}) as follows:
\begin{equation}
	v_z=-i \frac{k c_s^2}{\omega^2} A_m J_m(nr) \exp{i \left(kz+m\theta-\omega \left(t-t_A(r)\right)\right)} . \label{eq:vz+ta}
\end{equation}

As the wave frequency is constrained by the observation, we can derive the wave numbers $k$ for each azimuthal mode $m$ from the dispersion relation of \citep{Edwin1983}
\begin{equation}
	\rho n_e \left(\omega^2-k^2 c_A^2 \right) \frac{K'_m(n_e R)}{K_m(n_e R)}=\rho_e n \left(\omega^2-k^2 c_{A,e}^2 \right) \frac{J'_m (n R)}{J_m(n R)} ,
\end{equation}
where the subscript $e$ represents the exterior of the flux tube, $K_m$ is the modified Bessel function of the second kind, $K'_m$ is its derivative and $R$ is the radius of the tube, which is 5$\arcsec$ in our case.
We take $\omega=2 \pi/160$~s$^{-1}$ from the observation, $c_s=9$~km~s$^{-1}$ from \citet{Maltby1986}, $c_A=300$~km~s$^{-1}$ from \citet{Khomenko2006}, $c_{s,e}=1.5c_s$ and $c_{A,e}=0.5 c_s$ from \citet{Edwin1983}, then the $k$ is approximately 4.36$\times 10^{-6}$ rad m$^{-1}$ for all azimuthal modes.

\begin{figure*}
	\plotone{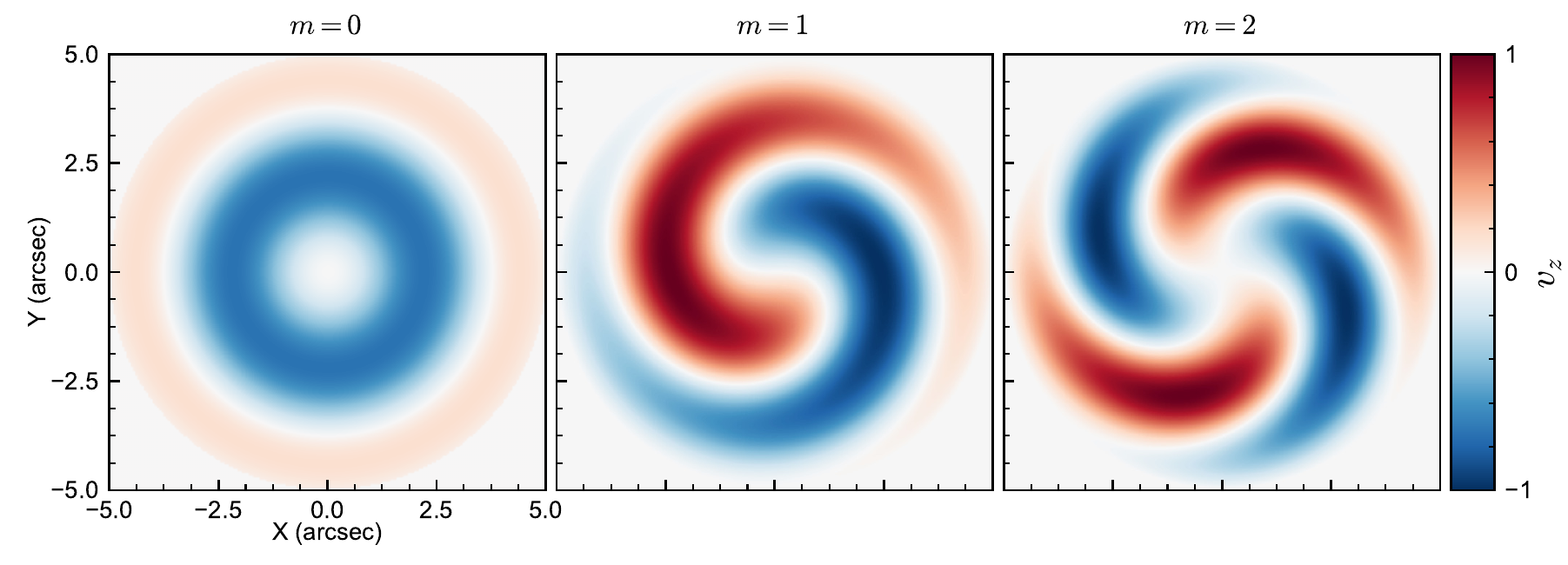}
	\caption{Snapshots of the simulated parallel velocity component for the azimuthal wave modes $m=0$, +1 and +2 at $t=0$ in $x-y$ plane.
		Speeds are normalized by the amplitude of each mode.
		The animation follows the azimuthal wave modes from $t=0\;\mbox{to}\;160\,\mbox{s}$.
		(An animation of this figure is available.)
		\label{fig:modes}}
\end{figure*}

Substituting these parameters into Equations (\ref{eq:vr}) and (\ref{eq:vz}), the ratio between the amplitudes of $v_z$ and $v_r$ is estimated as $v_z/v_r \sim 5\times10^3$ for all azimuthal modes.
It means that every azimuthal slow-body mode is predominantly longitudinal in the chromosphere.
Figure \ref{fig:modes} shows snapshots of $v_z$ for $m=0$, 1, and 2 modes in the $x-y$ plane with $d=1600$ km and $v_\mathrm{fast}=20$ km s$^{-1}$.
For the case of $m=0$, the ring-like pattern is generated, and this ring apparently propagates radially outward.
On the other hand, $m=+1$ and +2 modes produce apparently rotating patterns in the counterclockwise direction with one- and two-armed structures, respectively.
As the ring-like pattern of $m=0$ mode propagates radially, the power of this changes with time and radius, while the power of non-zero modes depends only on the radius because the patterns of these modes do not move out \textbf(see the online animated version of Figure~\ref{fig:modes}).

\begin{figure*}
	\plotone{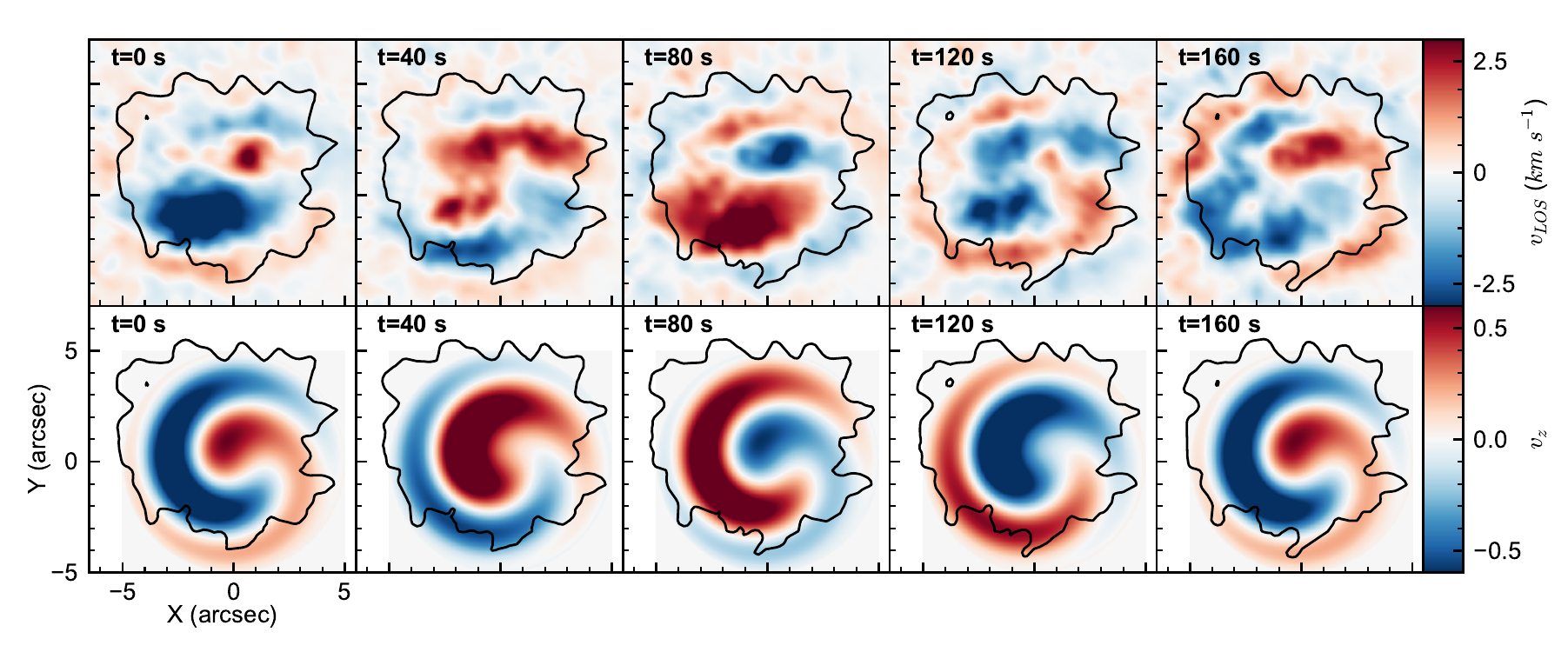}
	\caption{Time evolution of observed (top) and simulated (bottom) one-armed SWP from 17:17:20~UT to 17:20:00 UT.
	The observed Doppler maps are filtered in frequency bands from 5.5 to 9 mHz.
	The speeds in simulation are normalized by the maximum value.
	The boundary of the pore is shown by the solid line in both cases.
	The animation orients the observed data to the left and the simulation to the right; the observed data is presented in $20\,\mbox{s}$ increments while the simulation runs smoothly from $t=0\;\mbox{to}\;160\,\mbox{s}$.
	(An animation of this figure is available.)
	\label{fig:arm_1}}
\end{figure*}

To reproduce the observed one-armed spiraling pattern, we summed up perturbations with $m=0$ and $m=1$, which are the most powerful modes according to the Fourier analysis, with the amplitude ratio of $A_0/A_1=0.54$, the source depth of $d=1600$~km and averaged propagation speed of $v_\mathrm{fast}=20$~km~s$^{-1}$.
In addition, we introduce the reference time $t_0$ and reference angle $\theta_0$ terms to set the origin of the simulation, then the $t$ is replaced by $t-t_0$, and $\theta$ is substituted by $\theta-\theta_0$ in Equation~\ref{eq:vz+ta}.
Figure~\ref{fig:arm_1} indicates that the temporal evolution of the one-armed SWP from the observation (top) can be fairly well modeled by the simulation (bottom) with $t_0=-20$~s and $\theta_0=170^{\circ}$.
Like the observation, the simulation can make the one-armed SWP.
The red or blue arms abruptly change the trajectory to inward around $x=2\arcsec$, $y=1\arcsec$ in both the observation and the simulation.

\begin{figure*}
	\plotone{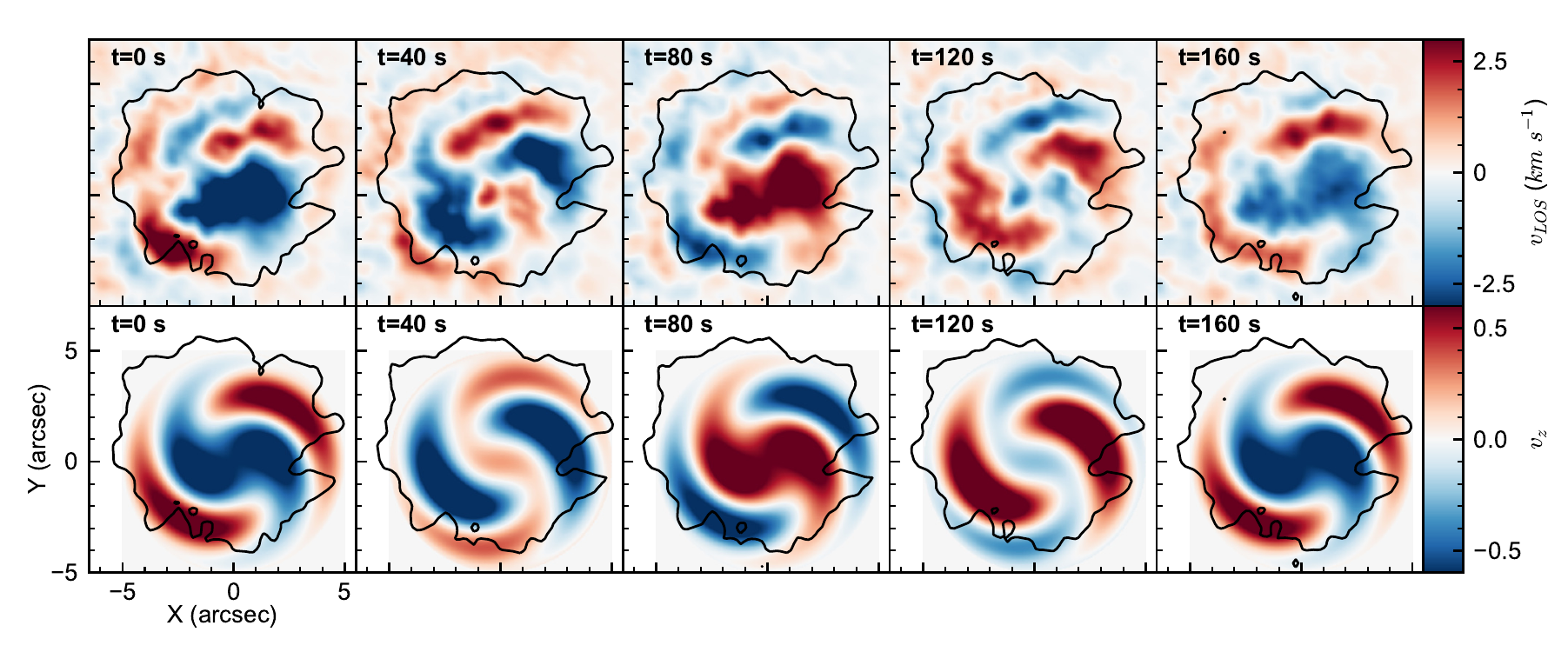}
	\caption{Similar to Figure~\ref{fig:arm_1}, but for the case of two-armed SWP from 17:44:07~UT to 17:46:47~UT.
	(An animation of this figure is available.)
	\label{fig:arm_2}}
\end{figure*}

We can successfully model the observed two-armed SWP as well.
Because the wave power is concentrated at $m=0$ and 2, we reproduce this pattern by summing up $v_z$ of $m=0$ and $m=2$ with the amplitude ratio of $A_0/A_2=0.54$, the reference time of $t_0=30$~s, and the reference angle of $\theta_0=30^{\circ}$.
In this simulation, the source is located at 1600~km below the $\beta=1$ layer and the averaged phase velocity is about 20~km~s$^{-1}$.
Figure~\ref{fig:arm_2} and associated animation represent the temporal evolution of the two-armed SWP.
The observation and simulation show quite similar two-armed spiraling features.
The two blue and red arms abruptly move inward around $x=-1\arcsec$, $y=2.\arcsec5$ and $x=1\arcsec$, $y=-2.\arcsec5$.

\section{Discussion} \label{sec:disc}

In this Letter, for the first time, we have presented a model that can explain the observed SWPs as slow magnetohydrodynamic (MHD) waves in an untwisted magnetic field.
In our model, the apparently rotating pattern is associated with the superposition of non-zero-$m$ azimuthal slow modes.
A non-zero-$m$ mode has a right-handed (left-handed) helical shaped wavefront for the case of positive (negative) $m$.
As this wave propagates upwardly along the straight field in a vertical magnetic flux tube, the wave pattern observed at some height shows an apparent rotation in the counterclockwise (clockwise) direction.
This kind of a rotating wave pattern was observed for the case of $m=1$ kink mode \citep{Jess2017}, and the related vortex dislocations were detected in a time-distance map along the slit placed in the center of the axis \citep{Lopez2016}.

The spiral structures and outward propagating wave patterns are formed by the internal driving sources, i.e. situated inside the magnetic flux tube forming the umbra, which are placed below the photosphere.
Beacause the wave propagates quasi-isotropically in the high-$\beta$ region, the longer the horizontal distance from the wave source to the observation point, the later the wave arrives.
The difference in the arrival times in the photosphere results in an apparent radially moving ring pattern in the case of $m=0$ (sausage) mode.
In non-zero-$m$ modes, the trailing spiral arm structures are formed because of the wave patterns rotate earlier as it is closer to the axis of the waveguiding flux tube. The number of arms depends on the absolute value of $m$. Thus, the observed apparent rotating spiral arms are not caused by the wave propagation in the azimuthal direction, but by the oblique, spiral-shaped wavefront of vertically propagating perturbations. 

Because of the abrupt spiraling motion of the one-armed spiral, \cite{Su2016} proposed that this pattern may be caused by the reflection at a light bridge.
In our case, however, there was not light bridge at all and, nevertheless, such SWPs were detected.
Our simulation clearly shows that the spiraling patterns are formed by the superposition of the wavefronts of an $m=0$ and a higher-$m$ modes. The one-armed SWP is generated by an $m=0$ sausage mode and an $m=1$ kink mode, and the two-armed SWP is formed by an $m=0$ sausage mode and an $m=2$ fluting mode.

We surmise that the driving source of a SWP may be associated with the downflows caused by the local magnetoconvection inside the sunspot. According to the 3D radiative MHD simulation of \citet{Kitiashvili2019}, acoustic waves can be generated by the converging downflows at 1.5 Mm beneath the surface inside a pore. This depth is very close to the depth of the source used for our model.
Furthermore, as there is no time lag between the two azimuthal modes in our simulation, it seems that these modes are excited simultaneously by the same driver.

We need to stress that the kink wave in a sunspot umbra or a pore considered here should not be confused with the kink waves studied in coronal loops.
In the loop, the kink mode is a transverse wave \citep{Aschwanden1999, Nakariakov1999}, while the sunspot kink mode considered here is a longitudinal wave associated with a slow magnetoacoustic wave \citep{Lopez2016, Jess2017}.
As a slow wave in a low-$\beta$ plasma, the kink wave in a sunspot is mainly characterized by parallel, field-aligned plasma flows. The radial flows, $v_r$, in this wave are quite small, because the $\omega^2-k^2 c_s^2$ factor in Equation~(\ref{eq:vr}) tends to zero as the phase speed is about the sound speed.
Another difference is connected with the wave polarization. Kink oscillations of coronal loops are usually linearly polarized, 
while the spiral wave structure in a sunspot requires the kink oscillation to be circularly polarized; i.e. the azimuthal wavenumber is $m=+1$ or $m=-1$. The sign is determined by the sense of rotation of the wavefront. 

Because the mechanism does not require additional assumptions such as the flux tube twisting or rotation, we expect that such SWPs may be generally detected in any sunspots.
As we accumulate the observation of those patterns, we can infer more physical parameters in sunspots such as propagating speed of fast wave and depth of the wave driving source.
Furthermore, those wave patterns can be considered as the evidence of the internal excitation of 3 minute oscillations in sunspots.
Further study of the SWPs may provide us with the clues to how magnetoconvection inside a sunspot generates such waves.

\acknowledgements
We greatly appreciate the referee's helpful and constructive suggestions and comments.
This work was supported by the National Research Foundation of Korea (NRF-2017R1A2B4004466).
V.M.N. acknowledges support from the STFC consolidated grant No. ST/P000320/1, and the BK21 plus program through the NRF funded by the Ministry of Education of Korea.

\bibliography{Juhyung}
\bibliographystyle{aasjournal}

\end{document}